\documentclass[aps,prl,twocolumn,superscriptaddress,notitlepage,10pt]{revtex4-1}
\usepackage{amsmath,amssymb,amsfonts}
\usepackage{graphicx}
\usepackage{natbib,hyperref}
\usepackage[dvipsnames]{xcolor}
\usepackage{comment}
\usepackage[overload]{empheq}
\usepackage{upgreek}
\usepackage{lmodern,pdftexcmds}

\newcommand{\eeq}{\end{equation}}
\newcommand{\beq}{\begin{equation}}
\newcommand{\beqa}{\begin{eqnarray}}
\newcommand{\eeqa}{\end{eqnarray}}

\begin{document}
\title{Anomalous dimensions of the Smoluchowski coagulation equation}

\author{Jens Eggers}
\affiliation{School of Mathematics, University of Bristol, Fry Building,
Woodland Road, Bristol BS8~1UG, United Kingdom}
\author{Marco Fontelos}
\affiliation{Instituto de Ciencias Matemáticas
(ICMAT, CSIC-UAM-UCM-UC3M), Campus de Cantoblanco 28049 Madrid, Spain}
\date{\today}

\begin{abstract}
The coagulation (or aggregation) equation was introduced by
Smoluchowski in 1916 to describe the clumping together of colloidal
particles through diffusion, but has been used in many different contexts
as diverse as physical chemistry, chemical engineering,
atmospheric physics, planetary science, and
economics. The effectiveness of clumping is described by a kernel
$K(x,y)$, which depends on the sizes of the colliding particles $x,y$. We
consider kernels $K = (xy)^{\gamma}$, but any homogeneous function can be
treated using our methods.
For sufficiently effective clumping $1 \ge \gamma > 1/2$,
the coagulation equation produces an infinitely large cluster in finite
time (a process known as the gel transition). Using a combination of
analytical methods and numerics, we calculate the anomalous scaling
dimensions of the main cluster growth, calling into question results
much used in the literature. Apart from the solution branch which
originates from the exactly solvable case $\gamma = 1$,
we find a new branch of solutions near $\gamma = 1/2$, which violates
scaling relations widely believed to hold universal.
\end{abstract}

\maketitle
Smoluchowski's equation \cite{S16} has its origin in physical
chemistry, but more generally furnishes a fundamental description
of the formation of larger objects by the aggregation of
smaller entities. It appears in many physical problems such as
planetesimal accumulation, mergers in dense clusters of stars, aerosol
coalescence in atmospheric physics, and polymerization and gelation
(see \cite{D72,Frie00,E86,JB87,L93,GMGO13}), but also in
chemical engineering \cite{GMGO13}, and the social sciences \cite{G16,BTHC18}
It describes
the evolution of the density $c(x,t)$ of particles of size $x$ at time
$t$, taking into account the formation of new clusters of size $x$ by
the aggregation of pairs of size $x-y$ and $y$ respectively, as well as the
disappearance of clusters of size $x$ forming a larger one:
\begin{gather}
c_{t}\left( x,t\right) =\frac{1}{2}\int_0^x K\left(x-y,y\right)
c\left( x-y,t\right) c\left( y,t\right) dy - \nonumber \\
c\left( x,t\right)\int_0^{\infty} K\left( x,y\right) c\left( y,t\right) dy.
\label{smoluchowski}
\end{gather}
Here the function $K(x,y)$ (known as the coagulation kernel)
describes the probability for two particles of sizes $x$ and $y$ to
stick together.

The behavior of solutions to \eqref{smoluchowski} depends crucially on the
degree of homogeneity of $K$. To explore this, here we restrict
ourselves to the class of models described by $K(x,y) = (xy)^{\gamma}$,
for which the degree is $2\gamma$. This kernel applies to branched polymers
with surface interactions \cite{HEZ83,E86}, and to fractal clusters more
generally \cite{Meak83,KBJ83}, but stands for a much broader
class of models whose asymptotic behavior for large cluster sizes scales
with an exponent $2\gamma$. One of the fundamental problems in the
field is to relate, by solving \eqref{smoluchowski}, $\gamma$
to the scaling exponent $\beta$ determining the {\it typical} size of
clusters, and the gel exponent $\tau$, giving the power-law size distribution
of clusters \cite{E86}. Thus by measuring $\beta$ or $\tau$, one is then
able to infer fundamental mechanisms of aggregation, in phenomena as diverse as
planatesimal formation \cite{GMGO13}, aerosol dynamics \cite{Fried66},
or pipeline fouling caused by asphaltenes \cite{VFMHHF17}.

Only for $\gamma = 1$ can \eqref{smoluchowski} be solved explicitly
\cite{MP06,MP08,EV10,HLN09}, for more general kernels studies have
relied on discrete particle simulations and ad-hoc scaling arguments
(see \cite{L03}, \cite{Lee} and references therein).
It is therefore of enormous importance to develop mathematical methods
able to provide novel information on the behavior of solutions to
(\ref{smoluchowski}).

For $1/2 < \gamma \le 1$, \eqref{smoluchowski} develops singularities
in finite time, such that, starting from an initial particle size
distribution $c(x,0)$ with all its moments
$M_{i}=\int_{0}^{\infty }x^{i}c(x,0)dx$ bounded,
there is a certain time $t_0$ such that all moments $M_{i}$ for
$i\geq i_{0}$ diverge (see \cite{EMP02} and references therein).
This phenomenon, which has the character of a phase
transition \cite{E86}, is called finite time gelation
(at a gelation time $t_0$), and indicates the aggregation of particles
in a single cluster of infinite mass. In practice, of course, the
singularity will be cut off by the finite size of the total
number of particles available. On the other hand, if
$\gamma \le 1/2$, solutions exist globally in time \cite{EM06}.

As in many other physical problems involving diverging quantities
(cf. \cite{EF09}), we assume that the
approach to the singularity is selfsimilar, of the form
$c(x,t) = t'^{\alpha}\psi(xt'^{\beta})$; here $t' = t_0 - t$ is the time
distance to the singularity.
However, this selfsimilar structure has so far only been
established for $\gamma = 1$, while for $\gamma < 1$ selfsimilar
solutions have not been determined explicitly. Discrete numerical
simulations (cf. \cite{MP04}, \cite{Lee}) appear to show
{\it selfsimilarity of the second kind} \cite{B96},
for which similarity exponents cannot be determined from
dimensional considerations, or from symmetry arguments. This fact was
proven in \cite{BF14} for $1-\gamma$ sufficiently small, without calculating
the exponents explicitly.

Inserting the similarity form into \eqref{smoluchowski} and balancing
powers of $t'$, one finds $\alpha -1=2\alpha -(2\gamma +1)\beta$,
so that $\alpha =\beta (2\gamma +1)-1$ and hence
\beq
c(x,t) = t'^{\beta(2\gamma+1)-1}\psi(x t'^{\beta}),
\label{sim}
\end{equation}
which will form the basis of our analysis; $s(t) = {\rm const} t'^{-\beta}$ is
known as the typical cluster size \cite{L03}. In addition, one obtains an
integral equation for the similarity profile $\psi(\xi)$.
The scaling form \eqref{sim} is also known as the
``self-preservation hypothesis''\cite{Frie00}; for example, using a rescaling
analogous
to \eqref{sim}, in Fig. 7.11 of \cite{Frie00} the distribution of
aerosol particles, taken from experiment, is collapsed onto a single
profile $\psi(\xi)$.

Imposing that the mass $M_{1}=\int_{0}^{\infty }xc(x,t)dx$
is conserved by selfsimilar solutions, one would obtain $\beta =1/(2\gamma-1)$.
However, this need not be the case; rather, mass only has to be conserved
by the full solution to \eqref{smoluchowski}, and not necessarily by an
{\it asymptotic} solution of the form \eqref{sim}. Below we will consider
such asymptotic solutions which do not conserve mass and for which the
cluster size diverges, and hence $\beta$ cannot be deduced from mass
conservation.

By considering the Laplace transform of $\psi$ \cite{BF14}, defined as
\beq
\Phi(\eta) =
\frac{1}{\beta}\int_0^{\infty} \left(1-e^{-\eta \xi}\right)\xi \psi(\xi) d\xi,
\label{L_sim}
\end{equation}
the equation for $\Phi$ becomes an equation similar
to those of non-local transport equations treated by us \cite{EF19},
and for which we have developed efficient numerical treatments.
The behavior of $\Phi(\eta)$ for large arguments
represents the distribution of small clusters; if
$\psi(\xi) \propto \xi^{-\tau}$ \cite{L03}, then
$\Phi(\eta)\propto \eta^{\tau-2}$, where $\tau$ is known as the
(pre)gel exponent \cite{DE85}. This implies that as $t_0$ is approached,
the cluster size distribution function approaches a power law
$c(x,t_0) \propto x^{-\tau}$, which will once more link to the kernel's
exponent $\gamma$. Similar power law distributions occur in many other
fields, such as Zipf's law \cite{G16}, and have been proposed to understand
phenomena such as bank mergers \cite{BTHC18}.

To derive the integral equation for $\Phi(\eta)$ to be used in the
following, we start from (2.6) of \cite{BF14}, which, adopting the
notation of the present paper, reads
\beq
-\nu\Phi(\eta) + \eta\Phi'(\eta) =
\frac{1}{2}\frac{d }{d\eta}
\left[\frac{1}{\beta}
\int_0^{\infty}(1-e^{-\eta\xi})\xi^{\gamma}\psi(\xi) d\xi\right]^2,
\label{equ_prl}
\end{equation}
where $\nu =(1-(2\gamma -1)\beta )/\beta$.
Defining
\begin{equation}
F \equiv \frac{1}{\Gamma(1-\gamma)}
  \int_0^{\infty}
  \frac{\Phi(\eta+\zeta)-\Phi(\zeta)}{\zeta^{\gamma}} d\zeta
\label{F}
\end{equation}
and inserting \eqref{L_sim} into \eqref{F}, one obtains,
interchanging the order of integration, and performing the
integral over $\zeta$, that $F$ equals the square bracket in \eqref{equ_prl}.
This shows that
\begin{equation}
-\nu\Phi(\eta) + \eta\Phi'(\eta) =
\frac{1}{2}\frac{d F^2}{d\eta},
\label{psi_sim_e}
\end{equation}
with $F$ defined by \eqref{F}.

The left hand side
of \eqref{psi_sim_e} corresponds to the time derivative of the
Smoluchowski equation, which is expected to vanish for large $\eta$
in order for the solution to match to the ``background'' of the distribution
of small clusters. This matching condition \cite{EF_book} then implies that
$\Phi(\eta)\propto\eta^{\nu}$ for large $\eta$, which leads to the scaling
relation $\tau = 2 + \nu = 1 + 2\gamma - \sigma$ \cite{L03}, which
relates the gel exponent $\tau$ with the exponent $\sigma\equiv\beta^{-1}$,
determining the typical cluster size. We will see below that this is true
only for the ``lower'' branch of solutions, which grows out of the classical
case $\gamma=1$, but fails for the ``upper'' branch, first reported here.
\begin{figure}
  \centering
\includegraphics[width=0.9\hsize]{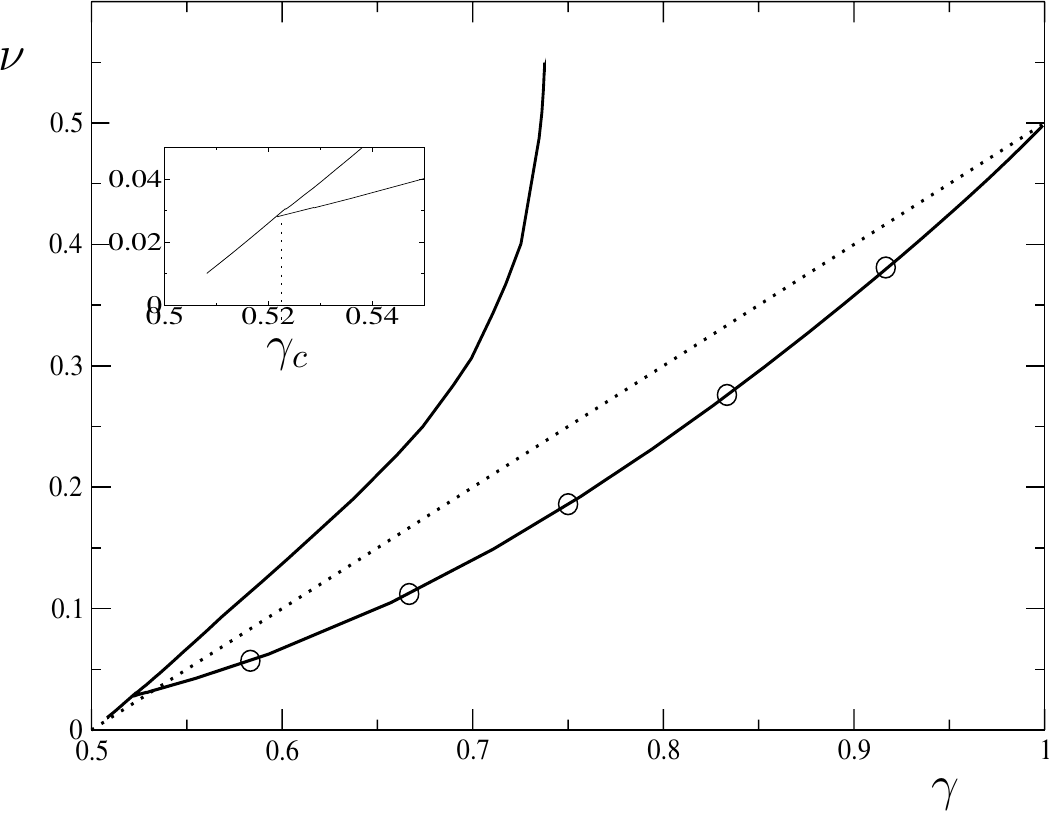}
\caption{The exponent $\nu$ for $\gamma$ between 1/2 and 1.
One branch (the lower branch) starts from $\gamma = 1$, $\nu=1/2$
on the right, and ends by intersecting another branch (the upper branch)
at $\gamma_c \approx 0.5214$.
The scaling $\nu = \gamma - 1/2$ \cite{L03} is shown as the dotted line.
The inset shows a detail of the bifurcation, where the two branches
meet at $\gamma_c$. The circles mark the particle-based simulations
of Lee \cite{L03}.
}
\label{fig:phase}
\end{figure}

We are looking for solutions to \eqref{psi_sim_e} with $\Phi(\eta)$
regular at the origin and
$\Phi(\eta)\approx A \eta^{\nu}$ for $\eta\rightarrow\infty$, where $A$ is
a constant to be found as part of the solution, along with $\nu$.
In the exactly solvable case
$\gamma = 1$, we have $F = \Phi(\eta)$, so \eqref{psi_sim_e} becomes
$\nu \Phi + \eta\Phi = \Phi\Phi'$, the same as for the kinematic
wave equation \cite{EF09,BF14}. This equation has an infinite sequence
of regular solutions $\eta = (1+j)\Phi/j + B \Phi^{1+j}$, where $j=1,2,\dots$,
and $B$ is an arbitrary constant. For the ``ground'' state $j=1$, the
cluster size exponent is $\beta = 2$, and $\nu = 1/2$. A sequence of
non-trivial solution branches, exhibiting anomalous scaling exponents,
emanate from each of these exact solutions; we will focus on the ground
state branch, which is expected to be attracting, while all other
branches are unstable.

To find solutions for $\gamma < 1$, \eqref{psi_sim_e} is solved
numerically, over $\eta\in [0,\infty[$, so it is useful to have a good
approximation of $\Phi(\eta)$ for large arguments. From
$\Phi(\eta)\approx A \eta^{\nu}$ it follows that
${\displaystyle F \approx A (\Gamma(\gamma-1-\nu)/\Gamma(-\nu))
\eta^{\nu+1-\gamma}}\equiv A C_{\nu}\eta^{\nu+1-\gamma}$,
from which one finds that $\Phi$ has the expansion
\beq
\Phi \approx A\eta^{\nu} + A^2\frac{\Gamma^2(\gamma-1-\nu)}
 {\Gamma^2(-\nu)}\frac{1-\gamma+\nu'}{2\gamma-1-\nu}
\eta^{1-2\gamma+2\nu} + \dots
\label{psi_next}
\end{equation}
for large $\eta$. As described in more detail in the supplementary material,
\eqref{psi_sim_e},\eqref{psi_next} are solved using a Newton method,
continuing from the ground state solution at $\gamma=1$.

\begin{figure}
\centering
\includegraphics[width=0.9\hsize]{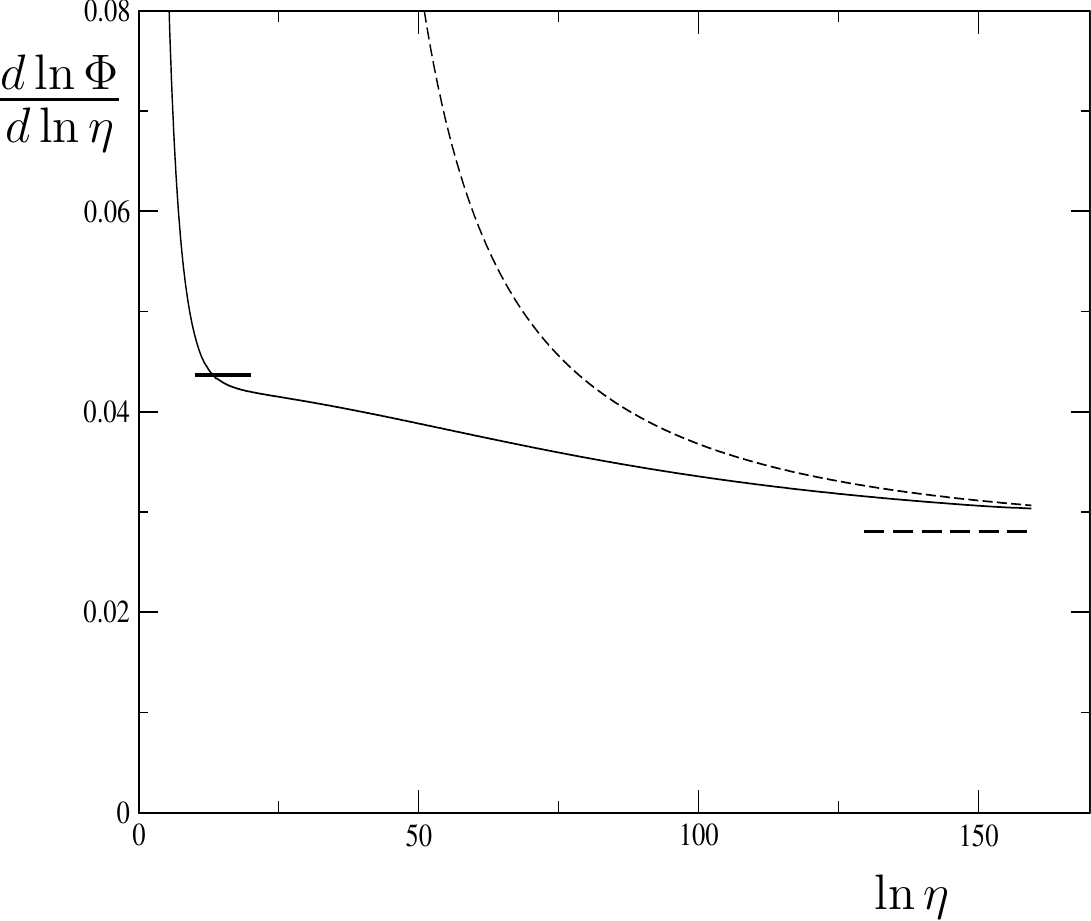}
\caption{The local exponent of $\Phi(\eta)$ over a large domain,
\eqref{psi_next} being used for $\eta > \eta_k = 10^{70}$,
for $\gamma = 0.521829$. The solid line is the
full solution on the lower branch, the dashed line is \eqref{psi_next}.
The solid horizontal line on the left is $\delta = 2\gamma-1 = 0.04366$,
the dashed horizontal line on the right is $\nu = 0.02808$.
        }
\label{fig:transition}
\end{figure}

The resulting values of $\nu$, which make up the ``lower branch'', are
shown in Fig.~\ref{fig:phase} as the lower solid line, emanating from
$\nu = 1/2$ for $\gamma = 1$. In a number of widely cited papers
\cite{DE85,HEZ83,ZEH83,LT82,L03}, it was proposed on the basis
of ad-hoc conditions on the behavior of the similarity equation for small
clusters, that $\sigma = (2\gamma -1)/2$ and $\nu = \gamma - 1/2$.
The latter is shown as the dotted line in Fig.~\ref{fig:phase},
clearly in strong
disagreement with the actual solution of \eqref{psi_sim_e}, as
anticipated in \cite{BF14}. Indeed, in \eqref{psi_sim_e} the behaviors
for small and for large clusters are in fact coupled, which leads to
{\it anomalous} scaling exponents \cite{B96,EF_book}, invalidating
a simple linear scaling. Our results also
agree very well with the numerics of \cite{Lee} (circles), obtained
using a particle-based description. For convenience, in the supplementary
material we also give an interpolation formula, which describes the solution
branch to three decimal places.

We have also calculated solution branches which emanate from the
higher-order solutions at $\gamma = 1$, which are known to be unstable
\cite{EF_book}. It is therefore likely that the entire higher-order branches
are unstable. Indeed, we have also solved the time-dependent evolution
equations in Laplace space for a particular value $\gamma = 0.7942$,
and found the solution to converge onto the stable ground state solution,
shown in Fig.~\ref{fig:phase}.

As $\gamma$ decreases toward 1/2, the correction exponent
$1-2\gamma+2\nu$ in \eqref{psi_next} becomes ever closer to
$\nu$, so a larger domain is needed to correctly describe the
asymptotics for large $\eta$, as seen in Fig.~\ref{fig:transition}.
The two exponents become identical for $\nu = 2\gamma - 1 \equiv\delta$,
which suggests the appearance of a new
branch of solutions, for which $\Phi(\eta) \sim \eta^\delta$, and which
we call the ``upper branch'', also shown in Fig.~\ref{fig:phase}.

To understand the transition between the two branches,
we write a formal solution of the similarity equation \eqref{psi_sim_e}:
\beq
\Phi(\eta) = \frac{F^2}{2\eta} +
\frac{1+\nu}{2}\eta^{\nu}
\int_0^{\eta} \frac{F^2}{\eta^{2+\nu}} d\eta.
\label{psi_formal}
\end{equation}
If the integral in \eqref{psi_formal} is convergent, then the second
term scales like $\eta^{\nu}$, and
\beq
A = \frac{1+\nu}{2}\int_0^{\infty}\frac{F^2(\eta')}{{\eta'}^{2+\nu}} d\eta';
\label{A}
\end{equation}
the first term in \eqref{psi_formal} is seen to be subdominant.

We anticipate a non-uniform convergence of the lower branch toward
the upper branch as $\gamma\rightarrow\gamma_c$. Let us assume
that as suggested by Fig.~\ref{fig:transition}, for
$1 \lesssim \eta \lesssim \eta_c$, $\Phi \sim \eta^{\delta}$,
while for $\eta_c \lesssim \eta \lesssim \infty$,
$\Phi \sim \eta^{\nu}$. Since the integral in \eqref{A} converges
for $\Phi \sim \eta^{\nu}$, this means that
$A \sim \eta_c^{\delta-\nu}$, where the exponent is positive on the lower
branch, so that as $\eta_c\rightarrow\infty$, the prefactor $A$ diverges.
Indeed, a best fit to the numerical data yields
$A \approx A_0 / (\gamma - 0.521)^\alpha$, with $\alpha = 0.348$ and
$A_0 = 0.44$, a blowup at $\gamma$ very close to the extrapolated value of
$\gamma  = \gamma_c = 0.5214$.

If on the other hand the integral in \eqref{psi_formal} is divergent,
both terms scale in the same way, and using the known asymptotic behavior
of $F$, balancing both sides yields
\beq
\Phi \approx \frac{2\gamma-1-\nu}{\gamma}
\frac{\Gamma^2(1-2\gamma)}{\Gamma^2(-\gamma)}\eta^{1-2\epsilon} \equiv
\bar{A} \eta^{2\gamma-1}
\label{psi_balance}
\end{equation}
to leading order as $\eta\rightarrow\infty$.

\begin{figure}
\centering
\includegraphics[width=0.9\hsize]{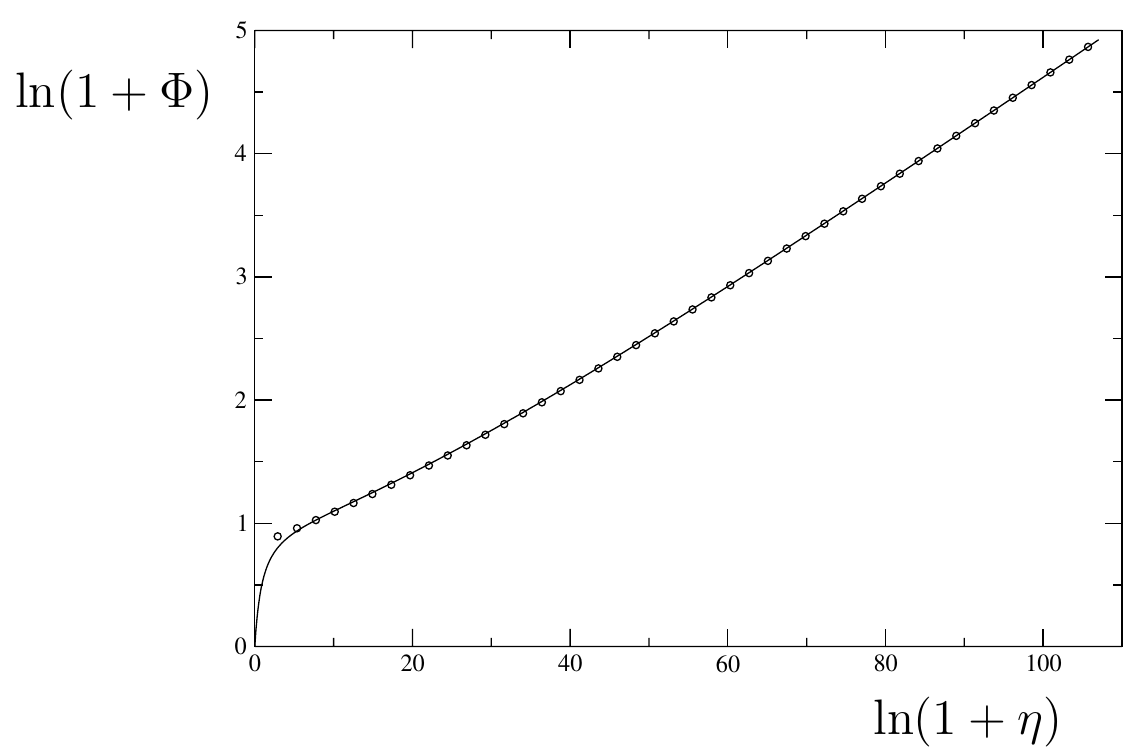}
\caption{
The profile $\Phi(\eta)$ on the upper branch, on a double logarithmic scale
(solid line), compared to the asymptotics \eqref{psi_balance}
(circles); $\gamma = 0.521829$, the same as in Fig.~\ref{fig:transition}.
                }
   \label{fig:upper}
\end{figure}

To find what we call the upper branch, we enforce
\eqref{psi_balance}  instead of \eqref{psi_next} for large $\eta$;
to account for the scale invariance of $\Phi$, we impose $\Phi'(0)=1$.
A typical profile on the upper branch is shown in Fig.~\ref{fig:upper},
using a logarithmic scale, except near the origin. The dashed line is
the expected asymptotics \eqref{psi_balance}, and $\gamma$ is the same
as in Fig.~\ref{fig:transition}, showing the lower branch. This demonstrates
that for a range of $\gamma$ values above $\gamma_c$, there are multiple
solutions. This is also clear from the phase diagram in Fig.~\ref{fig:phase},
where both branches are shown.
The lower branch ends at $\gamma_c$, where it meets the upper branch,
as seen in the inset. We show in the supplementary material that for
$\gamma = 1/2$, the
scaling function $\Phi(\eta)$ behaves asymptotically like a logarithm.
Thus for $\gamma \lesssim 1/2$ one needs an ever larger computational
domain to describe the crossover between logarithm and power law, and
we are not able to continue the upper branch all the way to $\gamma=1/2$.

A more fundamental concern is that on the upper branch,
$\eta^{\nu} \equiv \eta^{\alpha/\beta}$ does not equal the asymptotic behavior
$\Phi \propto \eta^{\delta}$, represented by circles in
Fig.~\ref{fig:upper}. As a result,
the Laplace transform of the cluster size distribution
\[
\omega(\mu,t) \equiv
\int_0^{\infty} \left(1-e^{-\mu x}\right)x c(x,t) dx =
t'^{\alpha-2\beta}\beta\Phi\left(\frac{\mu}{t'^{\beta}}\right)
\]
behaves for large arguments like
$\omega(\mu,t) \approx \beta \bar{A} \mu^{\delta} t'^{-1}$,
which for $t'\rightarrow 0$ diverges, and hence does not match the expected
static distribution at large arguments. However, the outer solution
\beq
\omega(\mu,t) = t'^{\lambda}\varphi(\xi),
\quad \xi = \mu t'^{-(1+\lambda)/\delta},
\label{sim_out}
\end{equation}
succeeds in bridging this time dependence with a static outer
distribution; $\lambda$ is another anomalous exponent to be determined.

The similarity equation becomes
$-\lambda\varphi + (1+\lambda)\xi \varphi_{\xi}/\delta = F F'$,
which for $\xi\rightarrow 0$ has solutions of the form
$\varphi \approx \beta \bar{A} \xi^{\delta} + G\xi^{\alpha}$.
Linearizing around the small perturbation $G\xi^{\alpha}$,
one finds
\[
\left(\frac{1+\lambda}{\delta}-\lambda\right)\alpha =
\frac{2(1+\alpha)C_{\alpha}}
{(1+\delta)C_{\delta}}.
\]
For small $\delta,\lambda$, this simplifies to
\[
2(1+\alpha)\Gamma(1/2-\alpha) = (1+2\alpha)\Gamma(1-\alpha)\sqrt{\pi},
\]
whose dominant (smallest) solution is $\alpha = \alpha_0 = 1.30737\dots\;$.
The exponent $\lambda$ plays the role of a nonlinear eigenvalue,
which has to be found such that the asymptotics of $\varphi$ at infinity
are satisfied.
On the other hand, for large $\xi$ we have to require that
$\varphi(\xi) \rightarrow \Phi_0\xi^{\lambda\delta/(1+\lambda)}$.
This will ensure that for small cluster sizes, $\omega$ matches onto a
static cluster size distribution.

There are many directions in which to extend the present research.
First, it would be interesting to consider \eqref{smoluchowski}
for times after the singularity,
and establish postgel solutions, including the scaling relations they
satisfy. Second, we have not yet explored the range of exponents
$0 \le \gamma \le 1/2$, for which an infinitely large cluster is
formed in the limit $t\rightarrow\infty$ only. For example, it is not known
whether in this case anomalous dimensions once more appear, or whether
the explicit expressions for $\sigma$ and $\tau$ in terms of $\gamma$,
established in the literature \cite{L03}, hold.

Third, many studies
have looked at other types of homogeneous kernels \cite{E86}, such
as $K = x^{\mu}y^{\nu}$ or $K = x^{\lambda} + y^{\lambda}$. For example,
an important question is whether exponents only depend on the degree
of homogeneity $\lambda = \mu + \nu$, or whether they are more sensitive
to the structure of the kernel.
Indeed, our methodology extends to much more general kernels,
for example of the form $K(x,y)=\left(x^{\mu }y^{\nu }+x^{\nu }y^{\mu }\right)/2$
(with $\mu +\nu =2\gamma$). This would lead to the integro-differential
equation \eqref{psi_sim_e}, with the right hand side replaced by
${\displaystyle \frac{1}{2}\frac{d(F_{\mu }F_{\nu })}{d\eta }}$, where $F_{\mu}$
and $F_{\nu}$ are defined as in \eqref{F}, with $\gamma $ replaced by
$\mu$ and $\nu$, respectively.

In wave turbulence \cite{H1,ZF66,NR11,N11}, similar integral equations
arise, which have not been solved explicitly, as we do here. Instead,
the theory rests on scaling assumptions similar to those which in
coagulation theory were found to be invalid. Here a stationary turbulent
spectrum would correspond to postgel solutions, which evolve out
of the initial singularity \cite{N11}. It is therefore possible that a
more careful treatment of the integral equations of wave turbulence
yields anomalous dimensions, as has been conjectured \cite{NR11},
which would change the scaling exponents of (say) the velocity field of
the turbulence.

In conclusion, integro-differential equations represent an area where
some of today's most challenging unsolved problems in statistical
mechanics and in fluid dynamics come together. Using self-similar solutions
to the Smoluchowski equation, we showed that such integral equations
have many unexpected properties, which challenge long-held beliefs.

\end{document}